# Machine Learning Models in Stock Market Prediction

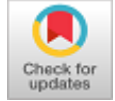

## Gurjeet Singh

*Abstract: The paper focuses on predicting the Nifty 50 Index by using 8 Supervised Machine Learning Models. The techniques used for empirical study are Adaptive Boost (AdaBoost), k-Nearest Neighbors (kNN), Linear Regression (LR), Artificial Neural Network (ANN), Random Forest (RF), Stochastic Gradient Descent (SGD), Support Vector Machine (SVM) and Decision Trees (DT). Experiments are based on historical data of Nifty 50 Index of Indian Stock Market from 22nd April, 1996 to 16th April, 2021, which is time series data of around 25 years. During the period there were 6220 trading days excluding all the non trading days. The entire trading dataset was divided into 4 subsets of different size-25% of entire data, 50% of entire data, 75% of entire data and entire data. Each subset was further divided into 2 parts-training data and testing data. After applying 3 tests- Test on Training Data, Test on Testing Data and Cross Validation Test on each subset, the prediction performance of the used models were compared and after comparison, very interesting results were found. The evaluation results indicate that Adaptive Boost, k- Nearest Neighbors, Random Forest and Decision Trees under performed with increase in the size of data set. Linear Regression and Artificial Neural Network shown almost similar prediction results among all the models but Artificial Neural Network took more time in training and validating the model. Thereafter Support Vector Machine performed better among rest of the models but with increase in the size of data set, Stochastic Gradient Descent performed better than Support Vector Machine.*

## I. INTRODUCTION

Stock Markets are the most popular financial market instrument. Perfect prediction of Stock Market Indices and stocks price is very difficult due to its highly dynamic nature. According to Abu-Mostafa & Atiya (1996), stock market prediction is considered as a very difficult task in financial time series prediction. According to Tan, Quek, & See (2007), stock market is affected by many macro economical factors such as political events, firm's policies, general economic conditions, investors' expectations, institutional investors' decisions, other stock market movements, and investors psychology etc. Lots of empirical studies dealt with the predicting stock index movement for the developed financial markets.

**Dr. Gurjeet Singh,** Associate Professor & Dean, Department of Lords School of Computer Applications & IT, Lords University, Alwar, Rajasthan, India. E-mail: research.gurjeet@gmail.com

Very less researches in the literature are found to predict the movements of index in the Indian Stock Market especially Nifty 50 Index.

My previous research works in the area of Stock Market (Singh & Nagar, 2012), (Nagar & Singh, 2012) and (Nagar & Issar, 2013) motivated me to do further study as still there are very limited researches have been done in the area of Stock Market.

The NIFTY 50 Index [39] consists of 50 stocks from 13 different sectors of the Indian economy.

ANN, LR, SGD, SVM, AdaBoost, RF, kNN and DT Machine Learning techniques are used for modeling and predicting the movements of the index.

The main objective of this empirical study is to explain and validate the expectedness of Stock Index movements using the above models and to make comparison of the performance of used techniques.

This research paper is divided into six sections. Section II provides the literature review of predicting Stock Market using Machine Learning. Section III explains the research data. Section IV tells about the Prediction Models used in this empirical study. Section V describes about the experiments and empirical results from the comparative analysis. And the last section, Section VI contains the concluding remarks.

## II. LITERATURE REVIEW

Using Artificial Neural Networks and Support Vector Machines, Y.Kara et al. (2011) attempted to predict the movements of stock price in the Istanbul Stock Exchange. They prepared 2 prediction models and compared their performances. The average prediction performance of the ANN model was 75.74% and SVM model was 71.52%. Alaa F et al. (2015) explored the use of Multiple Linear Regression and Support Vector Machine (SVM) to develop models for predicting the S&P 500 stock market index. 27 potential financial and economic variables which impact the stock movement were adopted to build a relationship between the stock index and these variables. The constructed SVM model with Radial Basis Function (RBF) kernel model provided good prediction capabilities with respect to the Regression and ANN models. Madge and Bhatt (2015) predicted the movements of stock price using SVM. 34 technology stocks and 4 parameters (technical indicators) were taken into consideration in their study. The work concluded that in the short-term predictions there was very low accuracy, on the other hand in the long-term prediction the prediction accuracy stood between 55 and 60%. Aparna Nayak et al. (2016) used historical and social media data.

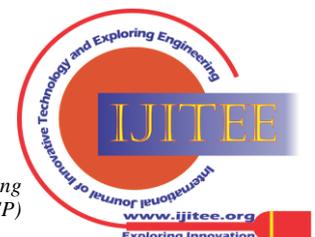

Prediction models were constructed using Decision Boosted Tree, Support Vector Machine and Logistic Regression to predict Indian Stock Market.

It was found that Decision Boosted Tree performed the best among the 3 models. Dash and Dash (2016) constructed a model which is called computational efficient functional link artificial neural network (CEFLANN) which generates the decisions of trading. CEFLANN provides profitable decision making with technical analysis rules. CEFLANN performance was compared with KNN, SVM, and DT, and CEFLANN provided the highest percentage of profit. The prediction of stock market behavior can depend on different types of ANNs because different type of ANNs can give different performance. It was observed that an input layer, hidden layer, and output layer combination can affect the performance of ANN. To get better performance, a right combination of layers is necessary. Sujin Pyo et al. (2017) employed an artificial neural networks (ANN), support vector machines (SVMs) with polynomial kernels, and radial basis function (RBF) kernels to predict the trend of the Korea Stock Price Index 200 (KOSPI 200) prices. In this study controversial issues were stated and hypotheses about the issues were tested. The results were inconsistent in comparison of the earlier research, which were assumed to have higher prediction rate of performance. Further it was proved by Google Trends that these were not effective factors in the prediction of the KOSPI 200 index prices. Next, the study also showed that ensemble methods do not improve the prediction accuracy. For prediction of the stock prices of having different capitalizations and different markets, Henrique et al. (2018) have used Support Vector Regression (SVR). The results obtained were indicating that SVR has a better predictive capability. Indu Kumar et al. (2018) have used Supervised Machine Learning Algorithms-Random Forest, SVM, KNN, Softmax Algorithm and Naive Bayes Algorithm for the prediction of stock price movements. The results of study revealed that in large dataset, Random Forest Algorithm had outperformed in comparison of all the other algorithms but with reduced size of the dataset (half of the original) Naïve Bayes Algorithm performed best accuracy wise. Next, decrease in the number of technical indicators will decrease the accuracy of each model in prediction of the stock market trends. Sheikh Mohammad Idrees et al. (2019) tried to analyze the time series data of Indian stock market. ARIMA model which is a statistical model was constructed to predict the Indian stock market volatility. Andrea Picasso et al. (2019) combined Technical and Fundamental analysis and constructed models to predict the market using Random Forest, Support Vector Machines, and a feed forward Neural Network applied to time series pre- diction and sentiment analysis. Irfan Ramzan Parray et al. (2020) attempted to predict NIFTY 50 index using SVM, Perceptron, and Logistic Regression. The SVM produced the best accuracy results among all algorithms.

## III. RESEARCH DATA

### Table 1. Data Sets

| | Data Set-1 | Data Set-2 | Data Set-3 | Data Set-4 |
|---|---|---|---|---|
| **Duration of Historical Data** | 22-Apr-96 to 8-Jun-99 | 22-Apr-96 to 10-Jul-02 | 22-Apr-96 to 19-Sep-08 | 22-Apr-96 to 16-Apr-21 |
| **Trading Days in Training Data** | 622 | 1244 | 2488 | 4976 |
| **Trading Days in Testing Data** | 156 | 311 | 622 | 1244 |
| **Total Trading Days in Dataset** | 778 | 1555 | 3110 | 6220 |
| **Percentage of Data Set (From Entire Data Set)** | 12.5% | 25% | 50% | 100% |

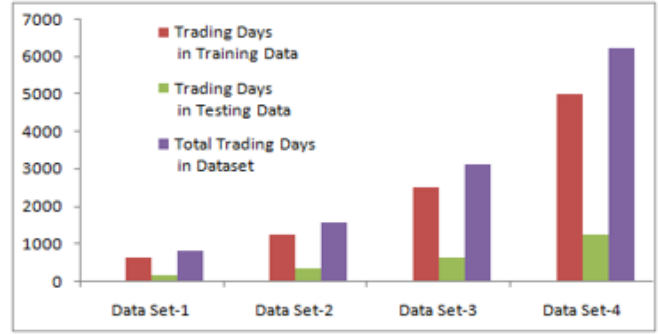

**Fig. 1. Data Sets of different size (12.5%, 25%, 50% and 100%)**

This section describes the research data and the selection of Data Sets. The research data used in this study is the direction of daily closing price movement in the Nifty 50 Index. The entire data set covers the period from April 22, 1996 to April 16, 2021. During the period, total numbers of trading days were 6220.

As shown in the Table 1 and Fig.1, four subsets were derived from the entire data set and each subset was further divided into 2 parts, Training Data Set (80% of the Data Set) and Testing Data Set (20% of the Data Set). The first subset is 12.5% of the entire data set and contains 778 trading day's data. This data set is called ''Data Set-1''. The Data Set-1 covers the period from April 22, 1996 to June 8, 1999, containing 622 trading day's data in Training Data Set and 156 trading day's data in Testing Data Set. The second subset is 25% of the entire data set and contains 1555 trading day's data. This data set is called ''Data Set-2''. The Data Set-2 covers the period from April 22, 1996 to July 10, 2002, containing 1244 trading day's data in Training Data Set and 311 trading day's data in Testing Data Set. The third subset is 50% of the entire data set and containing 3110 trading day's data.

This data set is called ''Data Set-3''. The Data Set-3 covers the period from April 22, 1996 to September 19, 2008, containing 2488 trading day's data in Training Data Set and 622 trading day's data in Testing Data Set. The fourth subset is 100% of the entire data set and contains 6220 trading day's data. This data set is called ''Data Set-4''. The Data Set-4 covers the period from April 22, 1996 to April 16, 2021, containing 4976 trading day's data in Training Data Set and 1244 trading day's data in Testing Data Set.

19

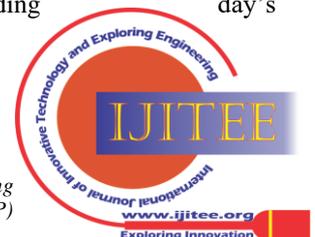

The performance comparison was performed on the Data Set-1, Data Set-2, Data Set-3 and Data Set-4 (entire Data Set) using ANN, LR, SGD, SVM, AdaBoost, RF, KNN and DT models. The results for each model and Data Set were compared to each other.

## IV. METHODOLOGY

### A. Adaptive Boosting (AdaBoost)

Boosting method is an ensemble method which is mostly used for predictions. This method is a group of algorithms which converts weak learners to a powerful learner. By using boosting weak learners are trained sequentially to correct its past performance. According to R.E. Schapire (2003) AdaBoost was the initial boosting algorithm which was successfully being used to meet the requirements of binary classification. AdaBoost algorithm is used to raise the performance of any machine learning algorithm, and it is mostly used with weak learners.

### B. K-Nearest Neighbors (kNN)

According to Duda et al. (2015), the k-Nearest Neighbors technique is a non-parametric technique which was used in the before 1970's in the applications of statistics (Franco-Loez et al., 2001). The basic theory behind kNN is that in the dataset of calibration, kNN is capable to find a group of k samples that are more nearest to unknown samples. The labels of unknown samples out of these k samples can be determined by finding the average of the response variables (Akbulut et al. and Wei et al., 2017). The k plays an important role in the performance of the kNN, because it is key tuning parameter of kNN (Qian et al., 2015). By using a bootstrap procedure, the parameter k can be determined.

### C. Linear Regression (LR)

Linear Regression is known as the linear relationship between the input data and the output data. The formula for a linear regression is as shown following:

$$y = c_0 + c_1x_1 + \ldots + c_nx_n$$

In the equation shown above there are n input variables; these are known as predictors or regressors and one output variable which is denoted by y (variable to be predicted). The constants $c_0, c_1 \ldots\ldots c_n$, are called the regression coefficients which is computed by principle of Least Square method. This is known Multiple Linear Regression because there are more than one predictor (Margaret H. Dunham, 2005). Regression Analysis is a methodology of statistics that is mostly used in numeric predictions (Jiawei Han and Micheline Kamber, 2010).

### D. Artificial Neural Network (ANN)

According to T. Subramaniam et al. (2010), first of all McCulloch and Pitts introduced Artificial Neural network (ANN) in 1943. Earlier ANN was widely used in words classification. ANN emulates the capabilities of human mind. In human mind neurons which are also called nerves cells, contact to each other through sending messages. ANN is highly capable and efficient data-driven model which is widely used in complex non linear behavior of data. ANN can be used in Hand Writing Recognition, Pattern Recognition, Face Identification, Text Translation, Medical Diagnosis, Speech Recognition, Credit Card Fraud Detection, Stock Market Prediction etc. Orange Data Mining Tool (Demsar, J. et al., 2013) provide Neural Network widget which uses Multi-Layer Perceptron (MLP) with back propagation, algorithm that can learn non-linear models as well as linear is used in this study. MLP is chosen as it is well suited in stock market prediction.

### E. Random Forest (RF)

According to Breiman et al. (2001), in Random Forest, 2 parameters are required: (i) ntree- the number of trees (ii) mtry- the number of features in each split. Many studies have declared that the suitable results could be obtained with the default parameters (Duro et al. and Immitzer et al. (2012); Liaw et al. (2002); Zhang et al. (2017)). According to Liaw et al. (2002), the more number of trees are giving more stable results. According to Breiman et al. (2001) use of more than the required number of trees may not be necessary, however this is not harmful to the model. According to Feng et al. (2015) with ntree = 200, accurate results can be obtained in Random Forest. Duro et al. (2012) states default value of mtry = √p, here p is denoted as number of predictors.

### F. Stochastic Gradient Descent (SGD)

According to L.Bottou (2010), in convex loss functions, Stochastic Gradient Descent is very much efficient approach for learning of unequal linear classifiers for example in SVM and logistic regression. SGD is being used for a long time in the Machine Learning society. In the context of large-scale learning, SGD is being widely used recently. Except large scale learning, SGD is being applied in sparse machine learning obstacle mostly in text classification and natural language processing.

### G. Support Vector Machine (SVM)

The Support Vector Machine algorithm can work well especially on the basis of classification. SVM is basically used due to its high generalization performance (Liu et al., 2017; Mohri et al., 2018). However the learning time of this algorithm is long, although more optimized results can be obtained by reducing the number of processes in learning. SVM is successful on large-volume data sets. SVM is used in prediction, image classification, medical diagnosis, text analytic, outlier detection etc.

### H. Decision Tree (DT)

According to J.-J.Sheu et al. (2016) Decision Tree is a data mining approach and it is based on the design of tree like structure. DT is a predictive modelling approach which is used in Statistics, Data Mining and Machine Learning. DT is mostly used in classification and regression based problems.

## V. EXPERIMENTS AND RESULTS

In this section, we conduct experimentation to evaluate the performance of Machine Learning models for Nifty 50 Index prediction on four different datasets.

20

The version 3.26.0 of Orange Data Mining Tool (Demsar, J. et al., 2013) was used in the experimentation and evaluation of the used Machine Learning Models in the study which can be downloaded from [37]. The Fig.2 shows the work flow of used Supervised Machine Learning Models in the study.

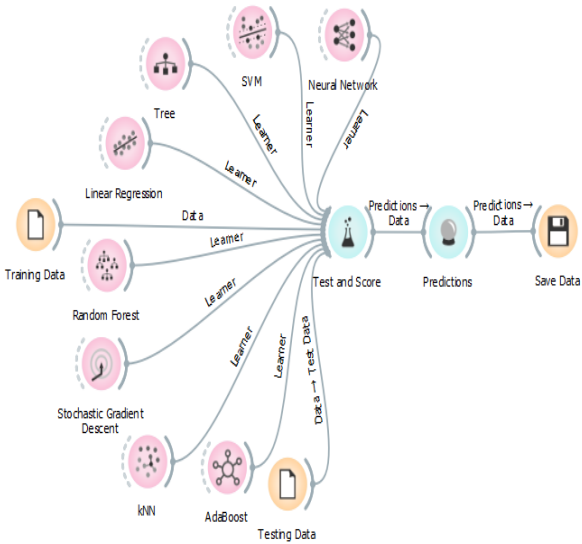

**Fig.2. Work Flow of Supervised Machine Learning Models**

### A. Features Extraction

Feature extraction is very important as wrong selection of features can result into wrong prediction. The historical data of Nifty 50 Index was downloaded from the website "niftyindices.com" [38] which is openly accessible, in CSV (Comma Separated Value) format. The available attributes in the historical data were S.No, Open, High, Low, Close and Date. S.No and Date attributes were skipped as these attributes are not important for prediction. As shown in the Fg.3, Open, High and Low attributes are selected as feature and Close attribute is selected as target because it is to be predicted.

| | Name | Type | Role |
|---|---|---|---|
| 1 | S.No | numeric | skip |
| 2 | Open | numeric | feature |
| 3 | High | numeric | feature |
| 4 | Low | numeric | feature |
| 5 | Close | numeric | target |
| 6 | Date | text | skip |

**Fig.3. Features Extraction**

### B. Parameter Settings for used Models

The setting of parameters for each model is important as wrong parameters setting can result into wrong output. The best combinations of parameters were set to get more accurate results. Figs. 4, 5, 6, 7, 8, 9, 10 and 11 are showing parameters setting of AdaBoost, kNN, LR, ANN, RF, SGD, SVM and DT respectively.

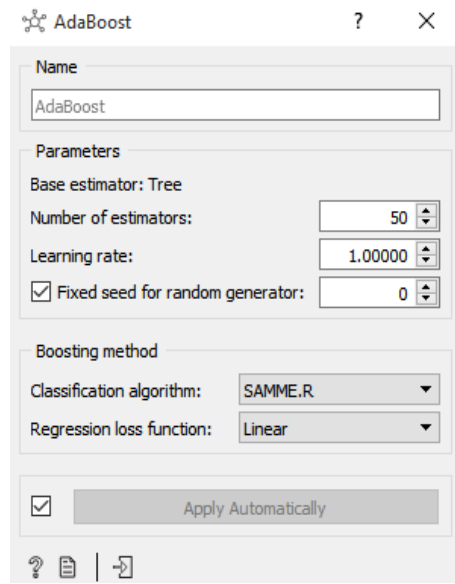

**Fig.4. AdaBoost Parameters**

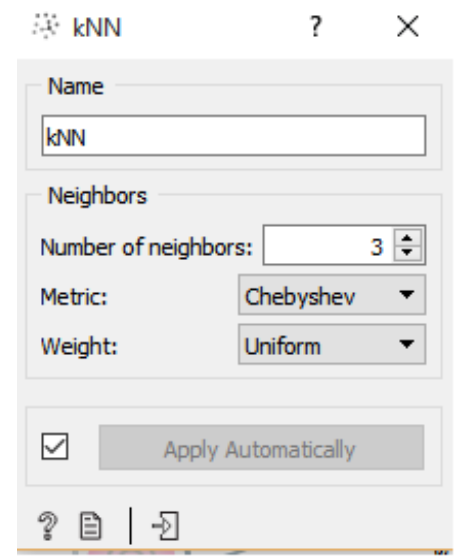

**Fig.5. kNN Parameters**

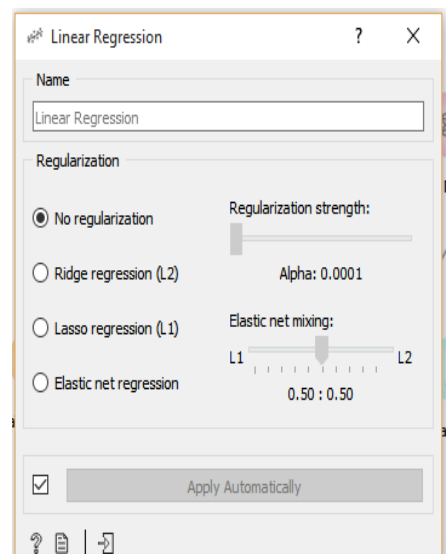

**Fig.6. Linear Regression Parameters**

21

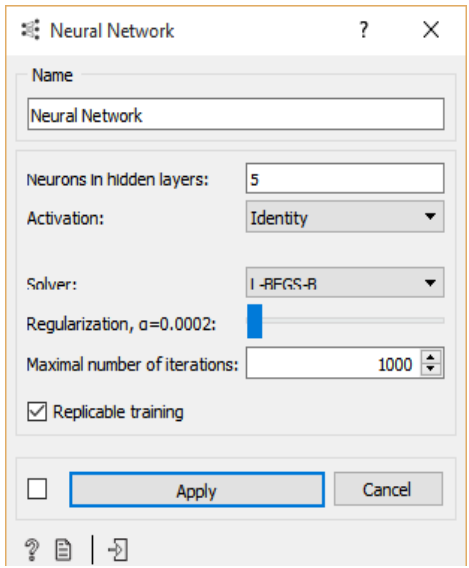

**Fig.7. Artificial Neural Network Parameters**

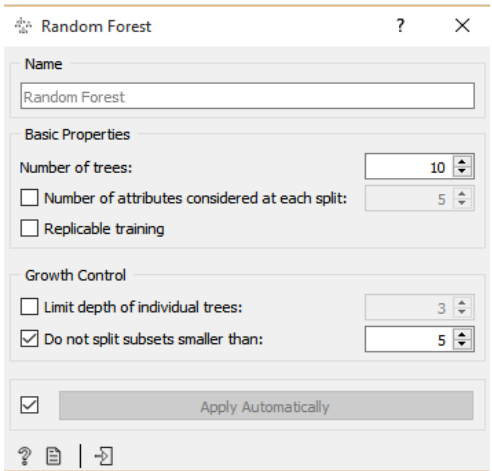

**Fig.8. Random Forest Parameters**

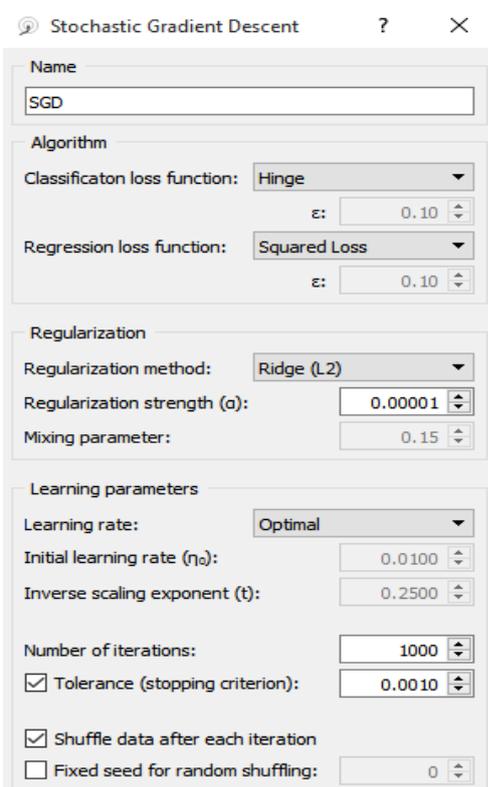

**Fig.9. Stochastic Gradient Parameters**

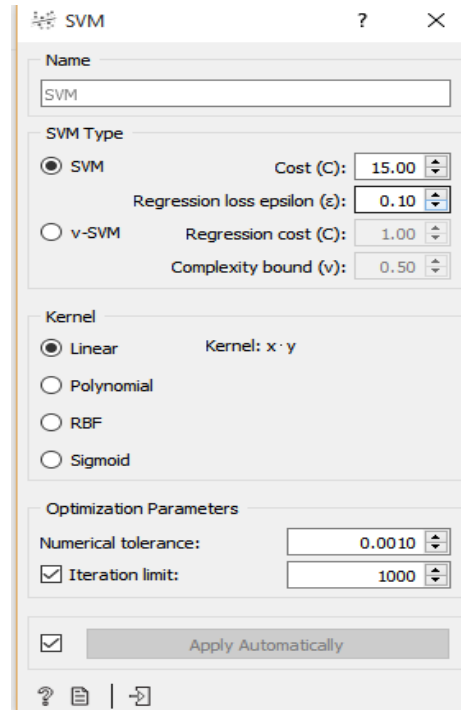

**Fig.10. Support Vector Machine Parameters**

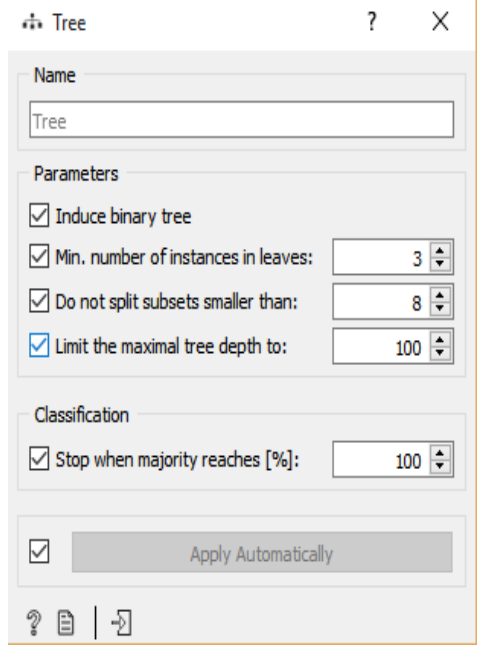

**Fig.11. Decision Tree Parameters**

### C. Testing Results on Data Set-1, Data Set-2, Data Set-3 and Data Set-4

In a dataset, a training dataset is implemented to build up a model, while a test (or validation) data set is used to validate the constructed model. Data points in the training datasets are excluded from the test dataset (validation). So, the training data is used to fit the model and testing data is used to test it. The ratio in each data set was taken 80% and 20% for Training Data and Testing Data respectively. Repeated k-fold cross-validation is used to improve the estimated performance of machine learning models.

**Table 2.  Cross Validation Test of Data Set-1**

| Rank | Model | Train time [s] | Test time [s] | MSE | RMSE | MAE | $R^2$ | CVRMSE |
|---|---|---|---|---|---|---|---|---|
| 1 | Linear Regression | 0.464 | 0.065 | 53.158 | 7.291 | 4.577 | 0.995 | 0.708 |
| 2 | ANN | 2.515 | 0.084 | 53.158 | 7.291 | 4.577 | 0.995 | 0.708 |
| 3 | SVM | 2.939 | 0.098 | 53.991 | 7.348 | 4.688 | 0.995 | 0.713 |
| 4 | SGD | 1.229 | 0.083 | 54.477 | 7.381 | 4.773 | 0.995 | 0.716 |
| 5 | kNN | 0.486 | 0.103 | 86.858 | 9.320 | 6.195 | 0.992 | 0.904 |
| 6 | AdaBoost | 10.653 | 0.532 | 92.244 | 9.604 | 6.327 | 0.992 | 0.932 |
| 7 | Random Forest | 2.548 | 0.156 | 95.146 | 9.754 | 6.614 | 0.992 | 0.947 |
| 8 | Decision Tree | 9.088 | 0.007 | 131.607 | 11.472 | 7.881 | 0.989 | 1.113 |

**Table 3. Test on Training Data of Data Set-1**

| Rank | Model | Train time [s] | Test time [s] | MSE | RMSE | MAE | $R^2$ | CVRMSE |
|---|---|---|---|---|---|---|---|---|
| 1 | AdaBoost | 1.296 | 0.082 | 0.517 | 0.719 | 0.235 | 1.000 | 0.070 |
| 2 | Random Forest | 0.294 | 0.019 | 20.645 | 4.544 | 3.079 | 0.998 | 0.441 |
| 3 | Decision Tree | 1.125 | 0.001 | 38.821 | 6.231 | 3.433 | 0.997 | 0.605 |
| 4 | kNN | 0.063 | 0.032 | 41.023 | 6.405 | 4.196 | 0.996 | 0.622 |
| 5 | Linear Regression | 0.053 | 0.016 | 51.801 | 7.197 | 4.536 | 0.995 | 0.698 |
| 6 | ANN | 0.395 | 0.020 | 51.801 | 7.197 | 4.536 | 0.995 | 0.698 |
| 7 | SVM | 0.382 | 0.042 | 52.367 | 7.236 | 4.633 | 0.995 | 0.702 |
| 8 | SGD | 0.161 | 0.009 | 52.759 | 7.264 | 4.706 | 0.995 | 0.705 |

**Table 4. Test on Test Data of Data Set-1**

| Rank | Model | Train time [s] | Test time [s] | MSE | RMSE | MAE | $R^2$ | CVRMSE |
|---|---|---|---|---|---|---|---|---|
| 1 | Linear Regression | 0.040 | 0.006 | 49.051 | 7.004 | 5.482 | 0.995 | 0.722 |
| 2 | ANN | 0.298 | 0.008 | 49.051 | 7.004 | 5.482 | 0.995 | 0.722 |
| 3 | SVM | 0.307 | 0.014 | 51.519 | 7.178 | 5.688 | 0.995 | 0.740 |
| 4 | SGD | 0.116 | 0.008 | 55.283 | 7.435 | 5.900 | 0.995 | 0.767 |
| 5 | kNN | 0.046 | 0.012 | 84.938 | 9.216 | 6.496 | 0.992 | 0.950 |
| 6 | Random Forest | 0.229 | 0.015 | 87.267 | 9.342 | 7.086 | 0.992 | 0.963 |
| 7 | AdaBoost | 0.984 | 0.046 | 98.286 | 9.914 | 7.316 | 0.991 | 1.022 |
| 8 | Decision Tree | 0.932 | 0.001 | 134.667 | 11.605 | 9.042 | 0.987 | 1.197 |

**Table 5. Cross Validation Test of Data Set-2**

| Rank | Model | Train time [s] | Test time [s] | MSE | RMSE | MAE | $R^2$ | CVRMSE |
|---|---|---|---|---|---|---|---|---|
| 1 | Linear Regression | 0.472 | 0.053 | 123.563 | 11.116 | 6.444 | 0.997 | 0.964 |
| 2 | ANN | 2.868 | 0.089 | 123.563 | 11.116 | 6.444 | 0.997 | 0.964 |
| 3 | SVM | 6.214 | 0.173 | 129.862 | 11.396 | 6.929 | 0.997 | 0.989 |
| 4 | SGD | 1.763 | 0.073 | 132.654 | 11.518 | 7.004 | 0.997 | 0.999 |
| 5 | kNN | 0.533 | 0.116 | 175.334 | 13.241 | 7.930 | 0.996 | 1.149 |
| 6 | Random Forest | 3.053 | 0.137 | 182.252 | 13.500 | 8.312 | 0.996 | 1.171 |
| 7 | AdaBoost | 14.946 | 0.520 | 192.136 | 13.861 | 8.350 | 0.996 | 1.203 |
| 8 | Decision Tree | 17.061 | 0.005 | 220.225 | 14.840 | 9.167 | 0.995 | 1.288 |

**Table 6. Test on Training Data of Data Set-2**

| Rank | Model | Train time [s] | Test time [s] | MSE | RMSE | MAE | $R^2$ | CVRMSE |
|---|---|---|---|---|---|---|---|---|
| 1 | AdaBoost | 1.813 | 0.107 | 0.948 | 0.974 | 0.360 | 1.000 | 0.084 |
| 2 | Random Forest | 0.364 | 0.022 | 47.379 | 6.883 | 3.933 | 0.999 | 0.597 |
| 3 | kNN | 0.058 | 0.043 | 78.852 | 8.880 | 5.288 | 0.998 | 0.770 |
| 4 | Decision Tree | 1.973 | 0.002 | 81.221 | 9.012 | 4.421 | 0.998 | 0.782 |
| 5 | Linear Regression | 0.054 | 0.006 | 122.394 | 11.063 | 6.428 | 0.997 | 0.960 |
| 6 | ANN | 0.459 | 0.021 | 122.394 | 11.063 | 6.428 | 0.997 | 0.960 |
| 7 | SVM | 0.721 | 0.124 | 124.990 | 11.180 | 6.699 | 0.997 | 0.970 |
| 8 | SGD | 0.214 | 0.008 | 126.222 | 11.235 | 6.761 | 0.997 | 0.975 |

**Table 7. Test on Testing Data of Data Set-2**

| Rank | Model | Train time [s] | Test time [s] | MSE | RMSE | MAE | $R^2$ | CVRMSE |
|---|---|---|---|---|---|---|---|---|
| 1 | Linear Regression | 0.046 | 0.007 | 30.471 | 5.520 | 4.134 | 0.993 | 0.512 |
| 2 | ANN | 0.341 | 0.008 | 30.471 | 5.520 | 4.134 | 0.993 | 0.512 |
| 3 | SVM | 0.685 | 0.034 | 31.231 | 5.588 | 4.274 | 0.993 | 0.518 |
| 4 | SGD | 0.179 | 0.008 | 32.043 | 5.661 | 4.387 | 0.993 | 0.525 |
| 5 | AdaBoost | 1.818 | 0.066 | 42.352 | 6.508 | 4.806 | 0.991 | 0.603 |
| 6 | Random Forest | 0.320 | 0.016 | 51.749 | 7.194 | 5.154 | 0.989 | 0.667 |
| 7 | Decision Tree | 1.824 | 0.001 | 62.872 | 7.929 | 5.889 | 0.986 | 0.735 |
| 8 | kNN | 0.052 | 0.015 | 69.306 | 8.325 | 5.031 | 0.985 | 0.772 |

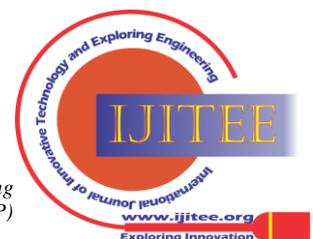

**Table 8. Cross Validation Test of Data Set-3**

| Rank | Model | Train time [s] | Test time [s] | MSE | RMSE | MAE | R² | CVRMSE |
|------|-------|------|------|------|------|------|------|------|
| 1 | Linear Regression | 0.721 | 0.057 | 147.960 | 12.164 | 6.101 | 0.999 | 0.893 |
| 2 | ANN | 6.591 | 0.101 | 147.960 | 12.164 | 6.101 | 0.999 | 0.893 |
| 3 | SGD | 4.054 | 0.081 | 161.002 | 12.689 | 6.933 | 0.999 | 0.932 |
| 4 | kNN | 0.783 | 0.153 | 231.892 | 15.228 | 8.010 | 0.999 | 1.118 |
| 5 | AdaBoost | 30.680 | 0.947 | 240.668 | 15.513 | 8.335 | 0.999 | 1.139 |
| 6 | Random Forest | 5.639 | 0.152 | 241.731 | 15.548 | 8.658 | 0.999 | 1.142 |
| 7 | Decision Tree | 38.699 | 0.011 | 296.749 | 17.226 | 9.890 | 0.999 | 1.265 |
| 8 | SVM | 14.939 | 0.451 | 307.789 | 17.544 | 12.945 | 0.999 | 1.288 |

**Table 9. Test on Training Data of Data Set-3**

| Rank | Model | Train time [s] | Test time [s] | MSE | RMSE | MAE | R² | CVRMSE |
|------|-------|------|------|------|------|------|------|------|
| 1 | AdaBoost | 3.039 | 0.302 | 0.887 | 0.942 | 0.358 | 1.000 | 0.069 |
| 2 | Random Forest | 0.515 | 0.026 | 56.254 | 7.500 | 3.994 | 1.000 | 0.551 |
| 3 | kNN | 0.087 | 0.078 | 107.462 | 10.366 | 5.355 | 1.000 | 0.761 |
| 4 | Decision Tree | 4.131 | 0.003 | 117.247 | 10.828 | 4.494 | 1.000 | 0.795 |
| 5 | Linear Regression | 0.051 | 0.007 | 148.125 | 12.171 | 6.089 | 0.999 | 0.894 |
| 6 | ANN | 0.811 | 0.017 | 148.125 | 12.171 | 6.089 | 0.999 | 0.894 |
| 7 | SGD | 0.430 | 0.009 | 157.518 | 12.551 | 6.772 | 0.999 | 0.922 |
| 8 | SVM | 1.615 | 0.367 | 361.298 | 19.008 | 14.896 | 0.999 | 1.396 |

**Table 10. Test on Testing Data of Data Set-3**

| Rank | Model | Train time [s] | Test time [s] | MSE | RMSE | MAE | R² | CVRMSE |
|------|-------|------|------|------|------|------|------|------|
| 1 | Linear Regression | 0.050 | 0.007 | 1329.200 | 36.458 | 23.923 | 0.998 | 0.850 |
| 2 | ANN | 0.501 | 0.008 | 1329.200 | 36.458 | 23.923 | 0.998 | 0.850 |
| 3 | SGD | 0.443 | 0.007 | 1396.104 | 37.364 | 26.271 | 0.998 | 0.871 |
| 4 | SVM | 1.558 | 0.089 | 1770.128 | 42.073 | 31.593 | 0.997 | 0.981 |
| 5 | AdaBoost | 3.135 | 0.170 | 1579476.492 | 1256.772 | 997.343 | -1.418 | 29.304 |
| 6 | Random Forest | 0.514 | 0.016 | 1642014.285 | 1281.411 | 1024.897 | -1.513 | 29.878 |
| 7 | kNN | 0.081 | 0.021 | 1653818.787 | 1286.009 | 1030.458 | -1.531 | 29.986 |
| 8 | Decision Tree | 3.741 | 0.001 | 1659037.594 | 1288.036 | 1033.104 | -1.539 | 30.033 |

**Table 11. Cross Validation Test of Data Set-4**

| Rank | Model | Train time [s] | Test time [s] | MSE | RMSE | MAE | R² | CVRMSE |
|------|-------|------|------|------|------|------|------|------|
| 1 | Linear Regression | 0.963 | 0.063 | 1767.328 | 42.040 | 13.638 | 1.000 | 1.223 |
| 2 | ANN | 9.136 | 0.084 | 1767.328 | 42.040 | 13.638 | 1.000 | 1.223 |
| 3 | SGD | 10.261 | 0.080 | 1879.098 | 43.349 | 16.298 | 1.000 | 1.261 |
| 4 | Random Forest | 9.196 | 0.186 | 2151.396 | 46.383 | 17.688 | 1.000 | 1.349 |
| 5 | AdaBoost | 64.507 | 1.637 | 2209.671 | 47.007 | 17.535 | 1.000 | 1.367 |
| 6 | kNN | 1.349 | 0.235 | 2292.070 | 47.876 | 16.926 | 1.000 | 1.392 |
| 7 | Decision Tree | 71.682 | 0.014 | 2328.751 | 48.257 | 20.030 | 1.000 | 1.404 |
| 8 | SVM | 28.303 | 0.710 | 5188.280 | 72.030 | 57.253 | 0.999 | 2.095 |

**Table 12. Test on Training Data of Data Set-4**

| Rank | Model | Train time [s] | Test time [s] | MSE | RMSE | MAE | R2 | CVRMSE |
|------|-------|------|------|------|------|------|------|------|
| 1 | AdaBoost | 7.840 | 0.462 | 3.561 | 1.887 | 0.780 | 1.000 | 0.055 |
| 2 | Random Forest | 0.966 | 0.038 | 443.957 | 21.070 | 8.215 | 1.000 | 0.613 |
| 3 | kNN | 0.157 | 0.136 | 1078.444 | 32.840 | 11.402 | 1.000 | 0.955 |
| 4 | Decision Tree | 7.838 | 0.008 | 1495.199 | 38.668 | 9.181 | 1.000 | 1.125 |
| 5 | Linear Regression | 0.096 | 0.006 | 1754.389 | 41.885 | 13.613 | 1.000 | 1.218 |
| 6 | ANN | 1.189 | 0.033 | 1754.389 | 41.885 | 13.613 | 1.000 | 1.218 |
| 7 | SGD | 0.980 | 0.009 | 1857.122 | 43.094 | 16.203 | 1.000 | 1.253 |
| 8 | SVM | 3.507 | 0.810 | 5129.207 | 71.618 | 56.908 | 0.999 | 2.083 |

**Table 13. Test on Testing Data of Data Set-4**

| Rank | Model | Train time [s] | Test time [s] | MSE | RMSE | MAE | R² | CVRMSE |
|------|-------|------|------|------|------|------|------|------|
| 1 | Linear Regression | 0.105 | 0.006 | 1358.994 | 36.865 | 25.722 | 0.999 | 0.348 |
| 2 | ANN | 0.796 | 0.008 | 1358.995 | 36.865 | 25.722 | 0.999 | 0.348 |
| 3 | SGD | 1.189 | 0.008 | 1802.542 | 42.456 | 29.093 | 0.999 | 0.400 |
| 4 | SVM | 3.123 | 0.175 | 4668.566 | 68.327 | 49.110 | 0.998 | 0.644 |
| 5 | AdaBoost | 7.222 | 0.206 | 5187963.087 | 2277.710 | 1737.004 | -0.930 | 21.474 |
| 6 | Random Forest | 1.000 | 0.017 | 5248175.973 | 2290.890 | 1750.752 | -0.952 | 21.598 |
| 7 | kNN | 0.142 | 0.039 | 5357927.984 | 2314.720 | 1772.651 | -0.993 | 21.823 |
| 8 | Decision Tree | 8.049 | 0.006 | 5379403.563 | 2319.354 | 1780.252 | -1.001 | 21.867 |

**D. Prediction Results for each used model with Actual Close Price of Nifty 50 Index**

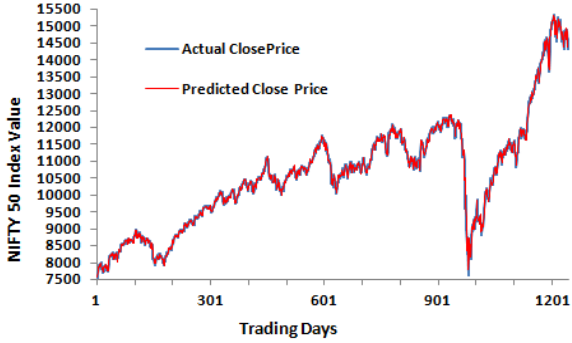

**Fig.12.Actual and Predicted Close Price of NIFTY Index using Linear Regression (LR) in Testing Data of DataSet-4**

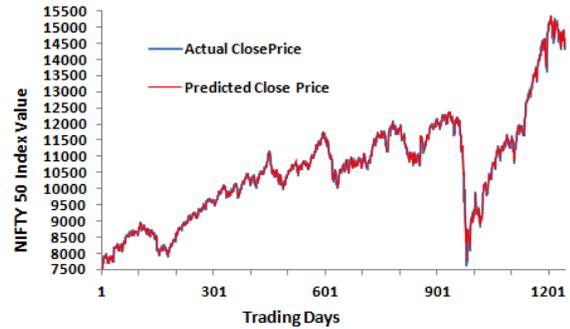

**Fig.13.Actual and Predicted Close Price of NIFTY Index using Artificial Neural Network (ANN) in Testing Data of Data Set-4**

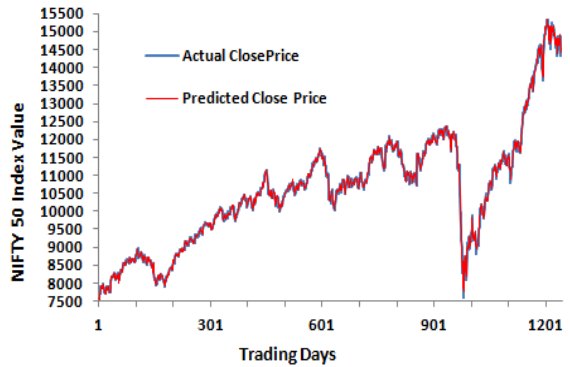

**Fig.14. Actual and Predicted Close Price of NIFTY Index using Stochastic Gradient Descent (SGD) in Testing Data of Data Set-4**

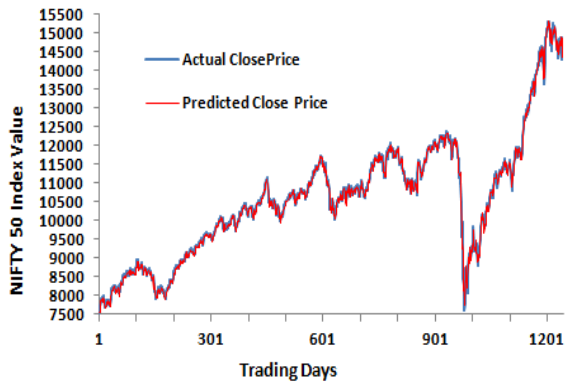

**Fig.15. Actual and Predicted Close Price of NIFTY Index using Support Vector Machine (SVM) in Testing Data of Data Set-4**

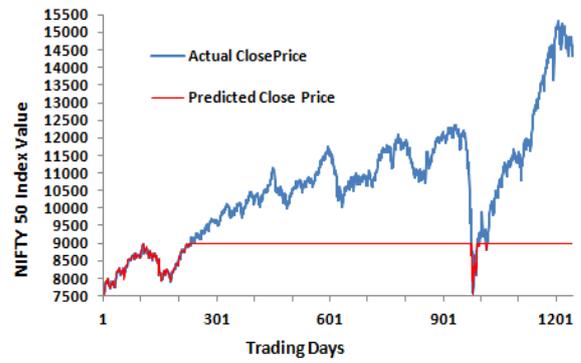

**Fig.16. Actual and Predicted Close Price of NIFTY Index using Adaptive Boost (AdaBoost) in Testing Data of Data Set-4**

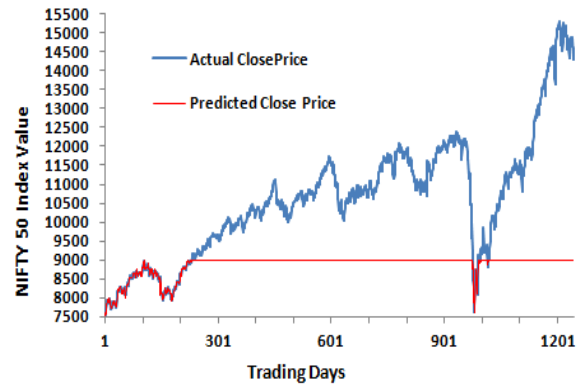

**Fig.17. Actual and Predicted Close Price of NIFTY Index using Random Forest (RF) in Testing Data of Data Set-4**

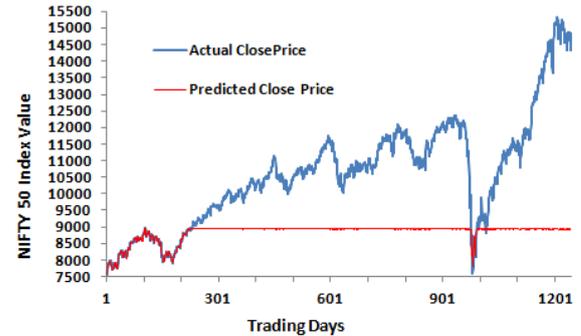

**Fig.18. Actual and Predicted Close Price of NIFTY Index using k-Nearest Neighbors (kNN) in Testing Data of Data Set-4**

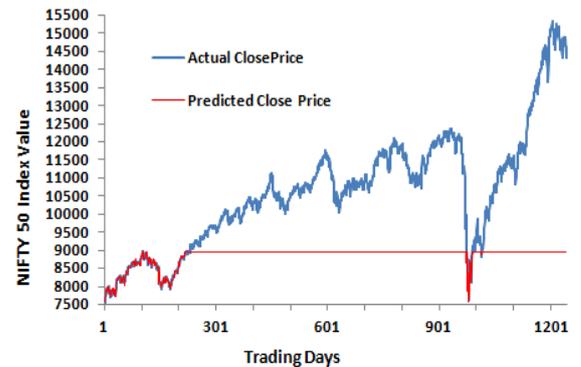

**Fig.19. Actual and Predicted Close Price of NIFTY Index using Decision Tree (DT) in Testing Data of Data Set-4**

### E. Discussion of Experiments and Results

Three tests were applied- Cross Validation Test, Test on Training Data and Test on Testing Data on each Data Set. These results are shown in the Tables 2, 3, 4, 5, 6, 7, 8, 9, 10, 11, 12 and 13 in the section V(C) Tables 2, 3 and 4 are showing the testing results of Data Set-1. Tables 5, 6 and 7 are showing the testing results of Data Set-2. Tables 8, 9 and 10 are showing the testing results of Data Set-3. Tables 11, 12 and 13 are showing the testing results of Data Set-4.

In the testing results the time taken to train the model and the time taken to test the model in seconds, also included to compare the execution time of the models.

The next measurements in the testing results are MSE, RMSE, MAE, $R^2$ and CVRMSE which stands for Mean Square Error, Root Mean Square Error, Mean Absolute Error, R-Square, Coefficient of the Variation of the Root Mean Square Error respectively. The lower value of MSE, RMSE, MAE and CVRMSE implies higher accuracy of a model. However, a higher value of R square is considered desirable.

Over fitting is identified by checking validation metrics such as accuracy and loss. The validation metrics are generally increased up to a certain point and after that start declining when the model is affected by over fitting. As a result, over fitting fails to fit additional data and this effect the accuracy of a predicting model. If the errors on the testing or validation dataset are higher than the errors on training dataset, it is said over fitting. Under fitting is identified when the model neither performs well with training data nor generalizes to test data. When a model performs well on Training Data as well as on Test Data then the model is a best fit or a good model.

Before comparing obtained testing results, sorting is done on RMSE variable, in ascending order for each table. It means the models having lower RMSE values are up in the tables in comparison to models having higher values of RMSE. After sorting the tables in ascending order on the basis of RMSE, rank is given to each model in increasing order as lower RMSE value will be having higher accuracy than the higher RMSE value.

In Data Set-1, from Tables 2 and 4, it is observed that LR and ANN results are almost similar but ANN takes more time in training and testing, so rank 1 is given to LR and rank 2 is given to ANN. Thereafter SVM and SGD performed well respectively among rest of the models, but SVM takes more time in training and testing in comparison to SGD. DT performed worst among all the models. Table 3 shows that AdaBoost, RF, DT and kNN performed well respectively in training even better than LR, ANN, SVM and SGD but Table 4 shows that these models did not performed as well in Validation Test.

In Data Set-2, from Tables 5 and 7, it is observed that once again LR and ANN results are almost similar and ranked $1^{st}$ and $2^{nd}$ respectively. SVM and SGD were ranked $3^{rd}$ and $4^{th}$ respectively. Table 5 shows that DT performed worst in Cross Validation Test and Table 7 shows that kNN performed worst in Validation Test.

Table 6 shows that AdaBoost, RF, kNN and DT performed well receptively in training even better than LR, ANN, SVM and SGD but Table 7 shows that these models did not performed as well in Validation Test.

In Data Set-3, from Tables 8 and 10, it is observed that this time also LR and ANN results are almost similar and ranked $1^{st}$ and $2^{nd}$ respectively. This time SGD performed better than SVM. SVM stood last in the Cross Validation Test. Table 9 shows that AdaBoost, RF, kNN and DT performed well receptively in training even better than LR, ANN, SGD and SVM but Table 10 shows that these models did not performed as well in Validation Test and obtained negative value of $R^2$. Negative value of $R^2$ on Test Data as shown in the Table 10 denotes the models AdaBoost, RF, kNN and DT are overfitted.

In Data Set-4, from Table 11 and 13, it is observed that LR and ANN results are almost similar and obtained rank 1 and 2 respectively. Thereafter SGD performed well compared to SVM. Table 12 shows that AdaBoost, RF, kNN and DT performed well respectively in training even better than LR, ANN, SGD and SVM but Table 13 shows that these models did not performed as well in Validation Test and obtained negative value of $R^2$. Negative value of $R^2$ on Test Data as shown in the Table 13 denotes the models AdaBoost, RF, kNN and DT are overfitted.

The above comparison made among all the models can also be understood by seeing the Figs. 12, 13, 14, 15, 16, 17, 18 and 19.

Fig.12. shows the Actual and Predicted Close Price of NIFTY 50 Index using Linear Regression (LR) in Testing Data of Data Set-4. Both the lines of actual and predicted close price of Nifty 50 Index are almost overlapping to each other; it means the prediction model is a good fit model for LR.

Fig.13. shows the Actual and Predicted Close Price of NIFTY 50 Index using Artificial Neural Network (ANN) in Testing Data of Data Set-4. Both the lines of actual and predicted close price of Nifty 50 Index are almost overlapping to each other; it means the prediction model is a good fit model for ANN.

Fig.14. shows the Actual and Predicted Close Price of NIFTY 50 Index using Stochastic Gradient Descent (SGD) in Testing Data of Data Set-4. Both the lines of actual and predicted close price of Nifty 50 Index are almost overlapping to each other; it means the prediction model is a good fit model for SGD.

Fig.15. shows the Actual and Predicted Close Price of NIFTY 50 Index using Support Vector Machine (SVM) in Testing Data of Data Set-4. Both the lines of actual and predicted close price of Nifty 50 Index are almost overlapping to each other; it means the prediction model is a good fit model for SVM. But comparing the Validation Test results of LR, ANN, SGD and SVM it is observed that LR and ANN performed better among all.

Thereafter SVM performed better than SGD in Data Set-1 and Data Set-2 but SGD performed better than SVM in Data Set-3 and Data Set-4. It indicates that with increase in the size of data set, SVM performance decreases.

Fig.16. shows the Actual and Predicted Close Price of NIFTY 50 Index using Adaptive Boosting (AdaBoost) in Testing Data of Data Set-4. The line of predicted close price of Nifty 50 Index overlaps the line of Actual Close price up to a certain point- 231 days out of 1244 days thereafter it remains constant as predicted close price in the output was also constant. It means the prediction model is an over fit model for AdaBoost.

Fig.17. shows the Actual and Predicted Close Price of NIFTY 50 Index using Random Forest (RF) in Testing Data of Data Set-4. The line of predicted close price of Nifty 50 Index overlaps the line of Actual Close price up to a certain point- 231 days out of 1244 days thereafter it remains constant as predicted close price in the output was also constant. It means the prediction model is an over fit model for Random Forest.

Fig.18. shows the Actual and Predicted Close Price of NIFTY 50 Index using k Nearest Neighbors (kNN) in Testing Data of Data Set-4. The line of predicted close price of Nifty 50 Index overlaps the line of Actual Close price up to a certain point- 231 days out of 1244 days thereafter it remains constant as predicted close price in the output was also constant. It means the prediction model is an over fit model for kNN.

Fig.19. shows the Actual and Predicted Close Price of NIFTY 50 Index using Decision Tree (DT) in Testing Data of Data Set-4. The line of predicted close price of Nifty 50 Index overlaps the line of Actual Close price up to a certain point- 218 days out of 1244 days thereafter it remains constant as predicted close price in the output was also constant. It means the prediction model is an over fit model for Decision Tree.

## VI. CONCLUSION

Prediction of the movements of the stock market index is very important for developing the effective market trading strategies. Financial decision to buy or sell an instrument may be made by the traders by choosing the effective predictive model. Successful prediction of Stock Market Index movements may be beneficial for investors. The tasks of predicting the movements of the Stock Market Index are highly complicated and very difficult. This empirical study attempted to predict the direction of Nifty 50 Index movement in the Indian Stock Market. Eight prediction models were constructed and their performances were compared on historical data from April 22, 1996 to April 16, 2021. Based on the experimental results obtained, some important conclusions can be drawn. Linear Regression and Artificial Neural Network performed almost equal performance in all the segments of different size of data. The reason behind good performance by Linear Regression competing to Artificial Neural Network was that regression method deals better with linear dependencies whereas neural networks can deal better with non linear dependencies. So if the data will be having some non linear dependencies, Neural Networks should perform better than regression. After LR and ANN, Support Vector Machine performed well but with increase in the size of data Stochastic Gradient Descent performed better than SVM. Thereafter ensemble learning methods of Decision Tree- AdaBoost and Random Forest performed better than kNN and Decision Tree with increase in the size of data.

In this empirical study eight Supervised Machine Learning Models were used, in future empirical study, more ensemble methods for Supervised Machine Learning Models can be taken. This empirical study used around 25 years historical data, which is good for machine learning because in such a long period many bull and bear phases of stock market were included.

27

## AUTHOR PROFILE

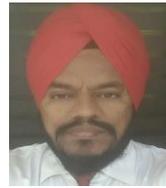

**Dr. Gurjeet Singh,** is working as Associate Professor & Dean in the department of Lords School of Computer Applications & IT, Lords University, Alwar, Rajasthan. He has rich experience of 22 years in teaching in the field of Computer Applications. He is awarded a Ph.D degree in Computer Science. His area of specialization is Data Mining. His interested areas of research are Outlier Detection, Data Management, Knowledge Discovery and Data Mining, Artificial Intelligence, Machine Learning and Deep Learning. He is also reviewer of reputed journals like IEEE Access. **Email:** research.gurjeet@gmail.com, gurjeet.singh@lordsuni.edu.in